\documentclass[aps,prb,superscriptaddress,twocolumn,floatfix,10pt]{revtex4-2}
\usepackage{amsmath}
\usepackage{amssymb}
\usepackage[mathscr]{euscript}
\usepackage{color}
\usepackage{hyperref} \hypersetup{colorlinks=true,citecolor=blue,linkcolor=blue,urlcolor=blue}
\usepackage{graphicx}
\usepackage{siunitx}
\usepackage{orcidlink}
\usepackage{chemmacros}
\usepackage{multirow}
\usepackage{upgreek}
\usepackage[normalem]{ulem}
\usepackage{siunitx}

\usepackage{float}

\makeatletter
\def\@hangfrom@section#1#2#3{\@hangfrom{#1#2}#3}
\def\@hangfroms@section#1#2{#1#2}
\renewcommand\frontmatter@abstractwidth{\dimexpr\textwidth\relax} \makeatother
\makeatletter

\usepackage{epstopdf}
\AppendGraphicsExtensions{.tif}

\def\AFLOW{{\small AFLOW}}

\def\POCC{{\small POCC}}

\def\VASP{{\small VASP}}
\def\DEED{{\small DEED}}
\def\QE{{\small QE}}
\def\HEC{{\small HEC}}
\def\HECs{{\small HECs}}
\def\PHEC{{\small PHEC}}
\def\PHECs{{\small PHECs}}

\def\EELS{{\small EELS}}
\def\FWHM{{\small FWHM}}
\def\RT{{\small RT}}
\def\HT{{\small HT}}
\def\DFT{{\small DFT}}
\def\TDDFPT{{\small TD-DFPT}}

\def\LFE{{\small LFE}}
\def\RPA{{\small RPA}}

\def\SEM{{\small SEM}}
\def\TEM{{\small TEM}}

\def\REELS{{\small REELS}}
\def\IR{{\small IR}}

\def\FAST{{\small FAST}}

\def\MBE{{\small MBE}}
\def\FCC{{\small FCC}}

\def\TM{{\small TM}}\def\TMs{{\small TM}s}
\def\ie{i.e., }
\def\eg{e.g., }

\definecolor{pranab_green}{rgb}{0,0.39,0}
\definecolor{pranab_red}{rgb}{0.85,0.23,0.11}
\definecolor{pranab_blue}{rgb}{0.13,0.18,0.40}

\def\SFigure{Supplementary Figure} \def\SFigures{Supplementary Figures}
 
\def\STable{Supplementary Table} 
\def\SIcolor{black}\def\Fcolor{black}

\newcommand\sgout{\bgroup\markoverwith{\textcolor{red}{\rule[0.5ex]{2pt}{0.4pt}}}\ULon}

\def\MEMS{\footnotesize Department of Mechanical Engineering and Materials Science, Duke University, Durham, NC 27708, USA}
\def\CEM{\footnotesize Center for Extreme Materials, Duke University, Durham, NC 27708, USA}
\def\ARLPSU{\footnotesize Applied Research Laboratory, The Pennsylvania State University, University Park, PA 16802, USA}
\def\MSPSU{\footnotesize Department of Materials Science and Engineering, The Pennsylvania State University, University Park, PA 16802, USA}
\def\MRIPSU{\footnotesize Materials Research Institute, The Pennsylvania State University, University Park, PA 16802, USA}
\def\MST{Department of Materials Science and Engineering, Missouri University of Science \& Technology, Rolla, MO 65409, USA}
\def\CNR{CNR-NANO Istituto Nanoscienze, Centro S3, I41125 Modena, Italy}

\begin{document}

\title{Variable-Temperature Plasmonic High-Entropy Carbides}

\author{Simon~Divilov\,\orcidlink{0000-0002-4185-6150}}\affiliation{\MEMS}\affiliation{\CEM}
\author{Sean~D.~Griesemer\,\orcidlink{0000-0001-5531-0725}}\affiliation{\MEMS}\affiliation{\CEM}
\author{Robert~C.~Koennecker\,\orcidlink{0009-0008-0810-373X}}\affiliation{\ARLPSU}\affiliation{\MSPSU}
\author{Michael~J.~Ammendola\,\orcidlink{0000-0003-2829-8331}}\affiliation{\ARLPSU}\affiliation{\MSPSU}
\author{Adam~C.~Zettel\,\orcidlink{0000-0003-1645-9476}}\affiliation{\MEMS}\affiliation{\CEM}
\author{Hagen~Eckert\,\orcidlink{0000-0003-4771-1435}}\affiliation{\MEMS}\affiliation{\CEM}
\author{Jeffrey~R.~Shallenberger\,\orcidlink{0000-0003-2590-1393}}\affiliation{\MRIPSU}
\author{Xiomara~Campilongo\,\orcidlink{0000-0001-6123-8117}}\affiliation{\CEM}
\author{William~G.~Fahrenholtz\,\orcidlink{0000-0002-8497-0092}}\affiliation{\MST}
\author{Arrigo~Calzolari\,\orcidlink{0000-0002-0244-7717}}
\email[]{arrigo.calzolari@nano.cnr.it}
\affiliation{\CNR}\affiliation{\MEMS}\affiliation{\CEM}
\author{Douglas~E.~Wolfe\,\orcidlink{0009-0004-5090-0290}}
\email[]{dew125@psu.edu}\affiliation{\MSPSU}
\author{Stefano~Curtarolo\,\orcidlink{0000-0003-0570-8238}}
\email[]{stefano@duke.edu}\affiliation{\MEMS}\affiliation{\CEM}

\begin{abstract}
\noindent
Effective thermal management at variable and extreme temperatures face limitations for the development of novel energy and aerospace applications.
Plasmonic approaches, shown to be capable of tailoring black-body emission, could be effective if materials with high-temperature and tunable plasmonic-resonance were available.
Here, we report a synergy between experimental and theoretical results proving that many high-entropy transition-metal carbides, consisting of four or more metals at equal molar ratio, have plasmonic resonance at room, high ($>$\SI{1000}{\degreeCelsius}) and variable temperatures. We also found that these high-entropy carbides can be tuned and show considerable plasmonic thermal cycling stability.
This paradigm-shift approach could prove quite advantageous as it facilitates the accelerated rational discovery and manufacturability of optically highly-optimized high-entropy carbides with ad-hoc properties.
\end{abstract}

\maketitle
\section*{\large Introduction} \label{sec:intro}

Thermal management  --- a challenge for efficiency and endurance of systems subjected to temperature variations~\cite{Shabany_heat_transfer_2009,Jafari_APSCI_2018} --- is crucial for extreme applications ($T$\SI{>1000}{\degreeCelsius}).
For example, understanding heat dissipation at {h}igh {t}emperature (\HT)~\cite{Kief_MRS_2018} and under intense electromagnetic and radiation fields~\cite{Wu_JOP_2012,Pillai_SSMSC_2010, Boriskina_MATT_2013,Cunha_AOM_2020,Zhao_NPHT_2021,Ahmadivand_MATT_2020,Jauffred_CHEMREV_2019} is key for the development of critical technologies, necessary for nuclear fusion reactors~\cite{McSherry_NNANO_2022} or aerospace applications~\cite{curtarolo:art202}.

In extreme environments, efforts to understand ---both theoretically and experimentally --- the physical mechanisms of heat/electrical conductivity, chemical degradation and optical response are limited in scope~\cite{Esser_JSR_2016,Dong_ENERGIES_2019}.
In fact, existing attempts to address thermal management at \HT, based on control of thermal conductivity demonstrate several limitations related to material endurance, structural integrity, and heat transport~\cite{Lv_ESS_2024}.
At extreme temperatures, thermal energy not only moves by lattice vibrations and charge carriers, but it is also emitted through the infrared part of the black-body radiation. Therefore, spectrum manipulation~\cite{Cuevas_ACSPHOT_2018} has opened a new avenue for thermal management. For example, hyperbolic metamaterials have been proposed to overcome the Planck black-body thermal-emission limit~\cite{Yang_NCOMM_2018,Guo_APL_2012,Nefedov_APL_2014} (so called super-Planckian~\cite{Xiao_NPHOTON_2022}), and Sm-doping has been used to change emissivity of \ch{ZrB2}/\ch{SiC} coatings for hypersonic flight.

Analogously, plasmonics would be promising since they are able to confine electromagnetic radiation to nanometer length-scales~\cite{Schuller_NMAT_2010,Tame_NPHYS_2013} and to amplify local electric fields~\cite{Beriki_NPHOT_2012}. For instance, plasmonic emitters~\cite{Costantini_PRAPPL_2015} provide tunable narrow band radiation, even narrower than the black-body radiation at the same temperature. Unfortunately, resonance tunability to manipulate different parts of the spectra is possible only at low temperature (e.g. by alloying Au and Ag). Thus, the unavailability of high temperature tunable plasmonic systems mandates novel materials that would  \textit{simultaneously} require:
{\bf i.} high structural resistance to deformations at \HT;
{\bf ii.} plasmonic activity in the near-IR/visible range under variable thermal conditions, where temperature may often change with large swings,
{\bf iii.} tunability, i.e. having a variety of resonance energies that can be modified by alloying; and
{\bf iv.} resistance to temperature cycles, i.e. restoring their initial plasmonic properties when the heating/cooling process is reversed.

Recently, in 2022, Calzolari et al.~\cite{curtarolo:art187} theoretically proposed that multifunctional {p}lasmonic {h}igh-{e}ntropy {c}arbides (\PHECs), a subset of the {h}igh-{e}ntropy {c}arbide (\HEC) realm~\cite{Oses_NatureReview_2020,unavoidable_disorder}, could combine superior thermal and mechanical stability with plasmon activity in the near-IR and visible range at \RT. With respect to  binary refractory ceramics (e.g. TiN, ZrN, TaC) -- which also stand out for their  high melting points ($>$\SI{3000}{\degreeCelsius}), impressive mechanical hardness~\cite{Guler_SCIENCE_2014, CedillosBarraza_HfTaC_SciRep_2016, Kumar_ACSPHOT_2016, Catellani_PRM_2017, Catellani_PRM_2020}, and plasmonic response in the visible light range at \HT\
~\cite{Tripura_JAP_2014, Reddy_ACSPHOT_2017, Krekeler_AOM_2021} --  the diverse composition of \PHECs\ would also enable tuning of critical properties, such as the plasmon resonance energy~\cite{Catellani_PRM_2020}.

In this article, we report a suite of experiments and calculations on \PHECs\ demonstrating: \textbf{i}. plasmon resonance in the near-IR and visible light ranges at \RT; \textbf{ii}. plasmonic character at \HT\ of at least \SI{1000}{\degreeCelsius}, with only minor changes with respect to the \RT; and \textbf{iii}. possibility to restore the initial plasmonic properties upon multiple thermal cycling from \RT\ to \HT.
We chose a set of transition-metal \PHECs\ by considering candidates with the highest likelihood of solid-solution functional synthesizability~\cite{DEED}, theoretical plasmonic intensity~\cite{curtarolo:art187} and structural stability~\cite{curtarolo:art140,curtarolo:art148}.
\PHEC\ ceramics were prepared by ball-milling mixtures of carbide precursors, followed by {f}ield-{a}ssisted {s}intering {t}echnology (\FAST) high-temperature sintering~\cite{Wolfe_JECERS_2022}.
Plasmonic properties were characterized by ellipsometry~\cite{Fujiwara_ellipsometry_2007} from \RT\ to \SI{1000}{\degreeCelsius}.
Reflective \EELS\ (\REELS) ~\cite{Wang_SURFSCI_1988} and \TEM-\EELS\ ~\cite{Egerton_eels_2011} were further used to confirm the optical results.
The experimental results were finally interpreted and analyzed through simulations from first-principles
with {d}ensity {f}unctional {t}heory (\DFT) that exploits the {p}artial {occ}upation (\POCC) disorder method~\cite{aflowPOCC} to evaluate the optical properties of \PHECs.

Here we identify a class of variable-temperature plasmonic \HECs\ that simultaneously satisfy the need for chemically tunable and structurally stable plasmonic materials that maintain an unaltered optical response over a wide temperature range (\SI{300}-\SI{1000}{\degreeCelsius}), and possibly beyond. Their high resistance to optical degradation upon extreme thermal treatments opens the way to unprecedented {\em variable temperature} plasmonic applications and impact manufacturability.
The combination of disorder thermodynamic modeling, first-principle calculations, systematic growth, and characterization of new \PHECs\ with customized optical properties, shows the benefit of our integrated approach.

\begin{table*}[t!]
\begin{tabular}{|c|c|cccccccc|cccc|}
\hline
\multicolumn{1}{|l|}{\multirow{2}{*}{}} & \multirow{2}{*}{\textbf{Composition}} & \multicolumn{8}{c|}{Ellipsometry} & \multicolumn{4}{c|}{Simulation} \\
\multicolumn{1}{|l|}{} & & \begin{tabular}[c]{@{}c@{}}$\mathit{E}_0$\\ \RT\\ {[}eV{]}\end{tabular} & \begin{tabular}[c]{@{}c@{}}$\mathit{E}_0$\\ \HT\\ {[}eV{]}\end{tabular} & \begin{tabular}[c]{@{}c@{}}$\mathit{E}_{\mathrm{peak}}$\\ \RT\\ {[}eV{]}\end{tabular} & \begin{tabular}[c]{@{}c@{}}$\mathit{E}_{\mathrm{peak}}$\\ \HT\\ {[}eV{]}\end{tabular} & \begin{tabular}[c]{@{}c@{}}\FWHM\\ \RT\\ {[}eV{]}\end{tabular} & \begin{tabular}[c]{@{}c@{}}\FWHM\\ \HT\\ {[}eV{]}\end{tabular} & \begin{tabular}[c]{@{}c@{}}$\mathit{h}_{\mathrm{peak}}$\\ \RT\\ {\small [a.u.]}\end{tabular} & \begin{tabular}[c]{@{}c@{}}$\mathit{h}_{\mathrm{peak}}$\\ \HT\\ {\small [a.u.]}\end{tabular} & \begin{tabular}[c]{@{}c@{}}$\mathit{E}_0$\\ {[}eV{]}\end{tabular} & \begin{tabular}[c]{@{}c@{}}$\mathit{E}_{\mathrm{peak}}$\\ {[}eV{]}\end{tabular} & \begin{tabular}[c]{@{}c@{}}\FWHM\\ {[}eV{]}\end{tabular} & \begin{tabular}[c]{@{}c@{}}$\mathit{h}_{\mathrm{peak}}$\\ {\small [a.u.]}\end{tabular} \\ \hline
\textbf{1} & \ch{HfNbTaZrC4} & 1.53 & 1.25 & 1.81 & 1.75 & 1.12 & 1.30 & 0.089 & 0.068 & 1.75 & 1.80 & 0.78 & 0.094 \\
\textbf{2} & \ch{HfNbTaTiZrC5} & 1.65 & 1.53 & 1.73 & 1.67 & 0.94 & 1.33 & 0.078 & 0.058 & 1.53 & 1.58 & 0.71 & 0.079 \\
\textbf{3} & \ch{HfNbTaWZrC5} & 2.05 & 1.83 & 2.34 & 2.27 & 1.50 & 1.73 & 0.048 & 0.038 & 1.87 & 2.08 & 1.42 & 0.064 \\
\textbf{4} & \ch{HfNbTaTiVC5} & 1.63 & 1.45 & 1.83 & 1.75 & 1.43 & 1.63 & 0.043 & 0.035 & 1.42 & 1.55 & 1.06 & 0.050 \\
\textbf{5} & \ch{HfNbTaTiWC5} & 1.86 & - & 2.27 & - & 1.66 & - & 0.055 & - & 1.81 & 1.98 & 1.28 & 0.060 \\
\textbf{6} & \ch{HfNbTaVZrC5} & 1.59 & 1.32 & 1.90 & 1.80 & 1.60 & 1.93 & 0.044 & 0.039 & 1.44 & 1.60 & 1.16 & 0.048 \\
\textbf{7} & \ch{HfNbTiVZrC5} & 1.37 & 1.16 & 1.64 & 1.51 & 1.39 & 1.87 & 0.040 & 0.038 & 1.19 & 1.31 & 0.95 & 0.042 \\
\textbf{8} & \ch{HfTaTiVZrC5} & 1.39 & 1.16 & 1.63 & 1.51 & 1.28 & 1.74 & 0.043 & 0.041 & 1.25 & 1.37 & 0.96 & 0.043 \\
\textbf{9} & \ch{HfTaTiWZrC5} & 1.81 & 1.61 & 2.00 & 1.92 & 1.28 & 1.54 & 0.051 & 0.038 & 1.64 & 1.78 & 1.13 & 0.058 \\
\textbf{10} & \ch{NbTaTiVWC5} & 2.17 & 2.04 & 2.29 & 2.20 & 2.08 & 1.99 & 0.027 & 0.020 & 1.75 & 1.84 & 1.05 & 0.043 \\
\textbf{11} & \ch{NbTaTiVZrC5} & 1.73 & 1.56 & 1.90 & 1.81 & 1.48 & 1.73 & 0.042 & 0.036 & 1.37 & 1.50 & 1.04 & 0.049 \\ \hline
\end{tabular}
\caption{\small \textbf{Properties from ellipsometry and calculations.} Optical properties of eleven {\PHEC} compositions from ellipsometry experiments and DFT simulations at room (\RT) and high temperatures (\HT=\SI{1000}{\degreeCelsius}, a.u. = arbitrary unit).}
\label{tab:peakparams}
\end{table*}

\section*{\large Results} \label{sec:results}

\noindent \textbf{Sample selection and synthesis} \\
\noindent We report new experimental results proving the existence of plasmon resonance in eleven \HECs;
the complete list and the compositions are summarized in Table \ref{tab:peakparams}.
A numerical label
(${{\bf 1}-{\bf 11}}$) is assigned to each system to simplify the notation.
System ${{\bf 1}}$ is the only 5-species carbide containing four transition metals (\TMs) plus carbon.
The remaining systems are 6-species \HECs\ with five \TMs.
Systems ${{\bf 1}-{\bf 3}}$ differ by a single species, and system ${{\bf 7}}$ is the only one not including Ta.

Individual carbide powder precursors
{\ch{HfC}, \ch{NbC}, \ch{TaC}, \ch{TiC}, \ch{VC}, \ch{WC}, \ch{ZrC}, and graphite} were blended together in mixtures according to the targeted candidate compositions.
\FAST\ sintering was used to densify the blended and dry milled powders.
More details about the synthesis of the compounds is discussed in the Methods Section.
All \HECs\ formed single-phase \FCC\ crystals at a soak temperature and time of at least \SI{2100}{\degreeCelsius} and 30 minutes, respectively; systems {\bf 6-8} and {\bf 11} required a higher temperature of \SI{2300}{\degreeCelsius} and a longer soak time of 60 minutes to reach miscibility.
The crystalline quality of the synthesized samples is demonstrated by the X-ray diffraction
spectra reported in
\textcolor{\SIcolor}{\SFigure\ 1},
where only small \ch{HfO2}-type oxide impurity peaks can be seen.
Subsequent analysis of the raw starting powder indicated that the oxide inclusions were the result of higher-than-expected oxide concentration in the precursors.
This was confirmed by {s}canning {e}lectron {m}icroscopy (\SEM)
images of the polished cross sections of \HECs\
(\textcolor{\SIcolor}{\SFigure\ 2})
which show the presence of oxide inclusions resulting from the impurities in the \ch{HfC} precursor.
Fortunately, \ch{HfO2} impurities are transparent dielectrics that are not expected to have plasmon response in the near-IR/visible range, nor to affect
the optical results of \HEC\ compounds.

Ten of the eleven compositions had been synthesized as \FCC\ single crystals before~\cite{Castle_2018_4metalC,
Zhou_2018_5metalC_oxidation, Yan_2018_5metalC_thermal_conductivity, curtarolo:art140, Chicardi_Carbides_CeramInt_2019, Chicardi_HECarbides_CeramInt_2020, curtarolo:art187}, whereas \textbf{3}-\ch{HfNbTaWZrC5} had not been previously synthesized.
All \HECs\ were computationally predicted to form single-phase \FCC, based on the use of the
{d}isordered {e}nthalpy-{e}ntropy {d}escriptor (\DEED)~\cite{DEED}. The successful synthesis of previously unrealized compound {\bf 3} provides additional validation of the \DEED\ method.
Details on sample preparation, experimental setups, and simulations can be found in the Methods Section.

\begin{figure*}[htb!]
\centerline{\includegraphics[width=0.95\textwidth]{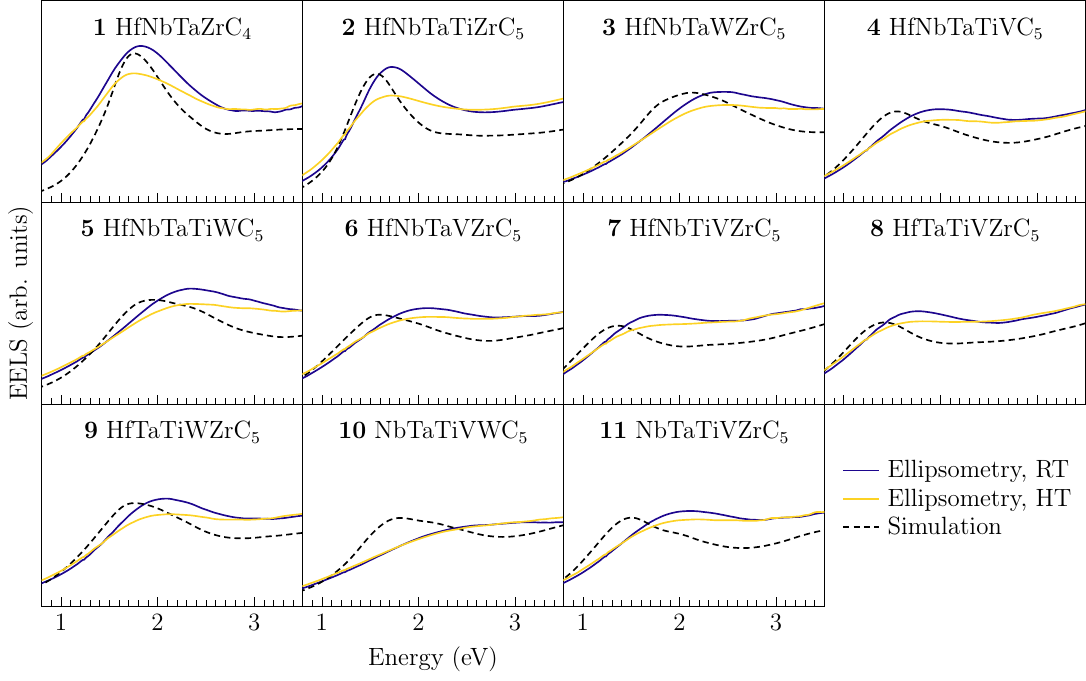}}
\caption{\small \textbf{Ellipsometry versus simulations.} Room temperature (\RT) and high temperature (\HT) \EELS\ spectra taken by ellipsometry, along with simulation results, for eleven \HEC\ compositions (\HT=\SI{1000}{\degreeCelsius}).
}
\label{fig:ellipsometry_ht}
\end{figure*}

\medskip
\noindent\textbf{Plasmonic resonance at room temperature} \\
\noindent The optical properties of \PHECs\ were measured by multi-technique spectroscopies.
Specifically, we used ellipsometry to measure the real and imaginary parts of the energy ($E$)-dependent complex dielectric function $\hat{\epsilon}({\bf q},E) = \epsilon_r({\bf q},E) + i\epsilon_i({\bf q},E)$, as well as the {e}lectron {e}nergy {l}oss {s}pectrum $\mathrm{EELS}({\bf q},E) = -\mathrm{Im}[\hat{\epsilon} ^{-1} ({\bf q},E)]$, in the optical limit of zero transferred momentum (${\bf q}={\bf 0}$)~\cite{Pines_ElementaryExcitations_1999}.
Plasmon excitations correspond to the poles of the complex dielectric function ($\hat{\epsilon}=0$).
As a consequence, the signature of a plasmon excitation is a peak in the \EELS, at energy $E_\textrm{peak}$, where the height $h_\textrm{peak}$ and {f}ull {w}idth at {h}alf {m}aximum (\FWHM) describe the intensity and the lifetime of the collective oscillations --- \ie, sharp (diffuse) peaks indicate long (short) lived plasmon excitations.
In this work, we focused on the low-energy part of the electromagnetic spectrum ($E<$~\SI{5.0}{eV}).
Finally, first principles simulations of the \POCC\ method have been carried out to compare and interpret the experimental results.
All technical details on both experiments and simulations can be found in the Methods Section.

In order to prove the robustness and the accuracy of the ellipsometry results, \REELS~\cite{Wang_SURFSCI_1988} and \TEM-\EELS~\cite{Egerton_eels_2011,curtarolo:art187} techniques were also used to measure the \EELS($E$) spectra of our \HECs.
Both ellipsometry and \REELS\ are surface sensitive techniques: ellipsometry analyzes the change in polarization of reflected light, while \REELS\ measures the energy loss of electrons reflected off the surface.
In comparison to \TEM-\EELS, \REELS\ has a much higher signal-to-noise ratio, allowing better determination of the peak position and distinguishing the characteristics of the plasmon peak with greater resolution.
\REELS\ has been used numerous times for plasmon detection in materials such as \ch{Si}~\cite{Yang_PRB_2019}, \ch{SiC}~\cite{Costantini_MAT_2023}, and graphene~\cite{Liu_PRB_2008}.
Experimental details on \REELS\ and \TEM-\EELS\
measurements are shown in
\textcolor{\SIcolor}{\SFigures\ 3 and 5.}

As a first step, we measure the plasmonic response of the \PHECs\ at \RT.
According to ellipsometry results (Figure \ref{fig:ellipsometry_ht}), all \PHECs\ exhibit a plasmon resonance in the range 1-\SI{3}{eV}; \ie spanning the near-\IR-visible portion of the electromagnetic spectrum, appealing for energy and telecommunication applications.

The agreement between ellipsometry (yellow line) and theoretical
simulations (dashed black line) is very good, corroborating the \REELS\ results
(\textcolor{\SIcolor}{\SFigure\ 4}).
Yet, ellipsometry and \REELS\ results differ slightly: \EELS\ peaks tend to be higher-energy, more intense, and wider than the respective \REELS\ peaks. Such small but systematic spectral differences can be attributed to
spot size and volume of analysis, which can be affected by changes in composition and microstructure
(\textcolor{\SIcolor}{\SFigure\ 3}).
Simulation results, while remarkably in line with both experimental sets of data, tend to agree more closely with ellipsometry.
More generally, the agreement between experiments and simulations is also remarkable as the simulations do not consider impurities, defects, or other microstructural features that may have effects on \PHECs\ at room temperature.
The excellent agreement between experiments and simulations is an a posteriori confirmation of both the crystalline quality of the synthesized samples and the accuracy of the \POCC\ method in modeling the optical properties of \PHECs.

In order to investigate the plasmonic nature of the \EELS\ peaks, we
analyze the real ($\epsilon_r$) and imaginary ($\epsilon_i$)
components of the dielectric function, resulting from the
ellipsometry measurements and corresponding to the spectra of
\textcolor{\Fcolor}{Figure~\ref{fig:ellipsometry_ht}}.
The data for \PHECs\ are summarized in
(\textcolor{\SIcolor}{\SFigure\ 6}).

The real part of the dielectric function is negative (metal-like) at energies of infrared and below, and crosses to positive (dielectric-like) in the near-\IR\ and visible light ranges at the ``crossover energy" $E_0$, so that $\epsilon_r(E_0)\equiv0$.
At the same energy, the $\epsilon_i$ reaches a local minimum near zero.
The condition $\hat{\epsilon}\simeq0$ corresponds to the excitation of a so called ``low-energy" or ``screened" plasmon resonance, which
involves the collective oscillation of a reduced fraction of the
charge density, the rest being effectively screened by interband transitions (the smaller the magnitude $\epsilon_i$, the longer the lifetime of the excitation).
Full charge density oscillations (\ie bulk plasmons) appear in \EELS\ spectra as intense peaks in the ultraviolet region ($E>$~\SI{20}{eV}).

\begin{figure}[ht!]
\centerline{\includegraphics[width=0.5\textwidth]{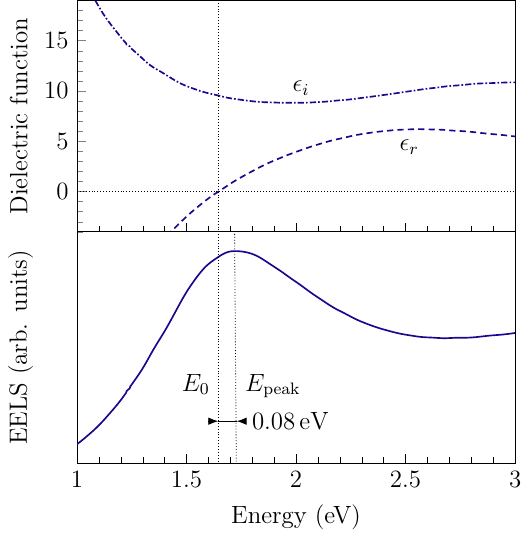}}
\caption{\small \textbf{Peak versus crossover energy.} Dielectric function and \EELS\ measured by ellipsometry at room temperature for \textbf{2}-\ch{HfNbTaTiZrC5}. }
\label{fig:HfNbTaTiZrC5_ellip}
\end{figure}

Close inspection of both ellipsometry and simulated spectra indicates that the crossover energy $E_0$ is very close to the measured peak maximum energy $E_{\mathrm{peak}}$, yet systematically smaller (within \SI{0.2}{eV}).
This is due to the non-ideal nature of the materials (\ie dissipation, losses). This idea is illustrated in \textcolor{\Fcolor}{Figure \ref{fig:HfNbTaTiZrC5_ellip}}.
In an ideal Drude electron gas, the whole complex dielectric function at crossover $\hat{\epsilon}(E_0)$ is zero, because of the lack of interband electronic dissipation leading to $\epsilon_i(E_0)=0$, and therefore $\epsilon_r(E_0)\equiv0$.
This corresponds to a divergence in the \EELS\ spectrum at $E_0=E_{\mathrm{peak}}$.
In dissipative systems, the condition $\epsilon_i\ne0$ removes the divergence,
while the positive slope $d\epsilon_i/dE >0$ for $E - E_0 \rightarrow 0^+$ is responsible for a positive contribution to the loss function that has a maximum at slightly higher energies, \ie $E_0 \lesssim E_{\mathrm{peak}}$.
The smaller the losses, the closer $E_0$ and $E_{\mathrm{peak}}$ are, like in our cases.
Therefore, we refer to $E_{\mathrm{peak}}$ as the plasmon energy of the system (see \textcolor{\SIcolor}{\SFigure\ 7} for further details).

Resonance energies listed in Table \ref{tab:peakparams} prove that \HECs\ have tunable plasmonic properties with respect to composition --- an advantage over other classes of plasmonic materials, such as noble metals and binary refractory ceramics.
In Ref.~\citenum{curtarolo:art187}, we discussed that the variation in $E_{\mathrm{peak}}$ between compositions is due to the balance among diverging effects induced by different transition metals. For instance, group-4 metals are responsible for red shifts of the plasmon energy, while group-6 metals cause a blue shift and broadening of the plasmon peak.

\begin{figure}[ht!]
\centerline{\includegraphics[width=0.45\textwidth]{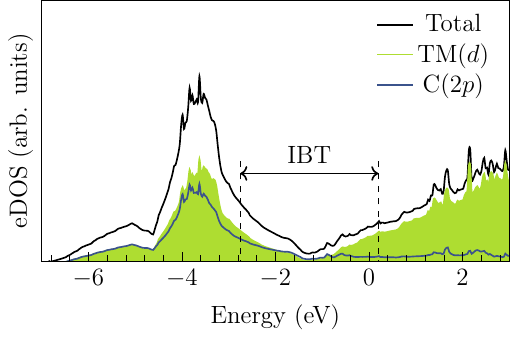}}
\caption{\small \textbf{Electronic density of states.} Example of eDOS for \textbf{1}-\ch{HfNbTaZrC4}, calculated for a single \POCC\ tile. Zero energy reference is set to Fermi energy.
Black line indicates the total eDOS, green area and blue line indicate the projected-eDOS on  transition metal (TM) $d$ states, and carbon $2p$ orbitals, respectively.
Black arrow pictorially represents an optically active interband transition (IBT) between occupied carbon $2p$ and empty transition metal (TM) $d$ bands. }
\label{fig:SI_DOS}
\end{figure}

For analyzing plasmon' lifetime, we fit the spectrum to a pseudo-Voigt function (combination of Gaussian and Lorentzian) with a linear background.
The resonance peak heights $h_\textrm{peak}$ and \FWHM\ are tabulated in Table~\ref{tab:peakparams}.
\EELS\ intensities of the plasmonic peaks
correlate to the density of states of TM
\textit{d}-orbitals ~\cite{curtarolo:art187,Dresselhaus_GroupTheory_2007}, crossing the Fermi energy, $E_F$
(an example of the density of states for \textbf{1}-\ch{HfNbTaZrC4} is shown in Figure~\ref{fig:SI_DOS}).
In particular, the $d$-like conduction states (above $E_F$)
are involved in optically active (i.e., dissipative) interband transitions from occupied \textit{p}-orbitals of carbon C$(2p)\rightarrow\mathrm{TM}(d)$~\cite{Kumar_ACSPHOT_2016}. While the carbon contribution is similar in all \PHECs, the energy distribution and the number of available empty $d$-states depend on the specific composition.
Therefore, the stronger and more numerous interband transitions lead to higher $\epsilon_i$; as a consequence, the \EELS\ peak is broader, and the lifetime of the plasmon excitation is shorter (\ie lossy plasmons).
The best 5-metal plasmonic candidate, having the highest $h_\textrm{peak}$ and lowest \FWHM\ within our set, is \textbf{2}-\ch{HfNbTaTiZrC5}.

If we consider this composition in nanostructured materials for applications utilizing surface plasmons, the quality factor of the local plasmonic resonance
$Q$~\cite{Wang_PRL_2006}
would be comparable to the best refractory carbide, \ch{TaC}.
All \HECs\ quality factors are listed in
\textcolor{\SIcolor}{\STable\ I.}
Notably, the parent binary rock-salt carbides (\eg \ch{TiC}, \ch{ZrC}, \ch{HfC}, \ch{VC}) are generally not plasmonic, except \ch{TaC}~\cite{curtarolo:art187}.
This indicates that the compositional disorder of \HECs\ is beneficial for optical properties and it can be exploited for optimizing plasmonic response.

\begin{figure}[b!]
 \centerline{\includegraphics[width=0.5\textwidth]{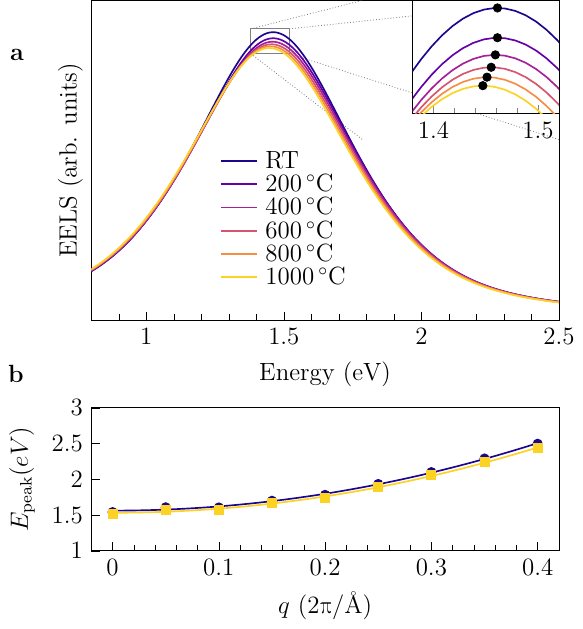}}
 \caption{\small \textbf{Simulated temperature dependence of the plasmonic properties of \ch{HfNbTaTiZrC5}.}
 {\bf a} Simulated \EELS\ at different temperatures. Inset shows the same spectra zoomed in at the peaks; the dots are the peak positions.
The discrepancy between the spectra is mostly due to neglecting the electron-phonon interactions in the calculations, as described in the main text.
 {\bf b} Simulated dispersion relation at room temperature (\RT) and \SI{1000}{\degreeCelsius}.
 }
 \label{fig:HfNbTaTiZrC5_Tdep}
\end{figure}

\medskip
\noindent \textbf{Plasmonic resonance at high temperature} \\
\noindent To verify that the \PHEC\ compositions remain plasmonic at \HT, we heat the samples to \SI{1000}{\degreeCelsius} at a rate of \SI{30}{\degreeCelsius/\minute}, followed by \SI{5}{\minute} of isothermal holding time, while performing the ellipsometry measurements.
High temperature \EELS\ spectra are shown in \textcolor{\Fcolor}{Figure
~\ref{fig:ellipsometry_ht}} (yellow lines), while the corresponding
dielectric function plots are reported in
\textcolor{\SIcolor}{\SFigure\ 6}.

The \HT\ values of $E_{\mathrm{peak}}$ and $E_0$ are summarized in Table \ref{tab:peakparams}.
At \HT, all of the \PHEC\ samples maintain their plasmonic character.
The spectra show that increasing temperature leads to only a slight red-shift ($\Delta E=$~0.04-\SI{0.2}{eV}) and weaker \EELS\ intensity.

In order to confirm the thermal stability of the plasmonic response of
\PHECs, we additionally performed \TEM-\EELS\ measurements of {\bf 2}-\ch{HfNbTaTiZrC5} at \RT\ and \SI{1200}{\degreeCelsius}, plotted in
\textcolor{\SIcolor}{\SFigure\ 5a}.
They show peak energies of $\sim$\SI{1.5}{eV}, very similar to the ones of simulation, ellipsometry, and \REELS.
While the \TEM\ peak is resolvable, the data is much noisier, due to the very small beam size of $\sim$\SI{1}{nm\squared} (comprising just a few unit cells), in contrast to the larger measurement regions of ellipsometry and \REELS\ ($\sim$\SI{2500}{\micro\meter\squared}).
Therefore, when taken at different spots on the same sample, \TEM-\EELS\ produces
variable peak parameters
(\textcolor{\SIcolor}{\SFigure\ 5b}),
whereas \REELS\ measurements at different spots are highly uniform
(\textcolor{\SIcolor}{\SFigure\ 3a}).
Overall, the \TEM-\EELS\ results confirm that {\bf 2}-\ch{HfNbTaTiZrC5}
is plasmonic of at least $T=\SI{1200}{\degreeCelsius}$, enlarging the thermal stability range spanned by ellipsometry.

The analysis of complex dielectric spectra at
\HT\
(\textcolor{\SIcolor}{\SFigure\ 6})
indicates that the reduction and the broadening in \EELS\ peak intensity observed in \textcolor{\Fcolor}{Figure~\ref{fig:ellipsometry_ht}} comes from a small decrease in the slope of $\epsilon_r$ and an increase in the local minimum value of $\epsilon_i$ at the energies near $E_\textrm{peak}$.
This is an indication of a larger dissipative scattering rate, which includes several electronic processes~\cite{Zollner_AOT_2022}.
For instance, the loss function can be expressed as $-\mathrm{Im}{\left[\hat{\epsilon}^{-1}(E_\textrm{peak})\right]} \approx E_\textrm{peak}/\gamma$, where $\gamma$ is the electron scattering rate, which includes the effects of thermal atomic displacement on the band structures,
and Fermi-Dirac broadening on the density of states around the Fermi level.
Electron-phonon interactions~\cite{Giustino_RMP_2017} further contribute to the changes of the \EELS\ peak.
They are described by non-Hermitian self-energy operators~\cite{Zollner_AOT_2022}, whose real part contributes to the red-shift of the plasmon energy,
while the imaginary part increases the decoherence of the electron density oscillations (\ie faster plasmon de-excitation) and reduces the \EELS\ peak intensity~\cite{Cohen_CondensedMatter_2016}.

Lattice thermal expansion~\cite{Baleva_JPCM_1990}, which reduces the effective free electron density, also contributes to the red-shift of $E_{\mathrm{peak}}$ with increasing temperature.
This manifests as an increase in $\epsilon_r$, such that the energy crossing (\ie the plasmon energy) occurs at smaller values~\cite{Reddy_ACSPHOT_2017}.
This result is also supported by first principles calculations in the temperature range 300-\SI{1000}{\degreeCelsius}.
In this case, the temperature is the
configurational temperature used to average the \POCC\ ensemble, and the calculations include the effect of thermal expansion~\cite{Yan_2018_5metalC_thermal_conductivity}.
The resulting spectra (\textcolor{\Fcolor}{Figure~\ref{fig:HfNbTaTiZrC5_Tdep}a}) agree qualitatively with our experiments (\textcolor{\Fcolor}{Figure~\ref{fig:ellipsometry_ht}}), reproducing the bathochromic shift of the plasmonic peaks with increasing temperature.
The decrease in the \EELS\ intensity is less severe in theoretical spectra than in experiments, because of lack of electron dissipative scattering effects in simulations that are known to increase with temperature~\cite{Kaveh_AP_1984}.

While the identification of the single scattering contributions at \HT\ is challenging, we indirectly investigated their effect on the plasmon lifetime by analyzing the \FWHM\ modifications in the ellipsometry spectra; the results are summarized in Table~\ref{tab:peakparams}.
We observe that the ranking of compositions by \EELS\ intensities is almost the same at both \RT\ and \HT\ with a Pearson correlation coefficient of 0.97.
While \textbf{2}-\ch{HfNbTaTiZrC5} is still the best material in our
set, its $Q$ value is not very high
(\textcolor{\SIcolor}{see \STable\ 1}),
yet still good enough to support surface plasmon devices, especially in view of the existing $Q$-enhancement techniques~\cite{Tandon_CHEMMAT_2019}.

\begin{figure*}[ht!]
 \includegraphics[width=1.0\textwidth]{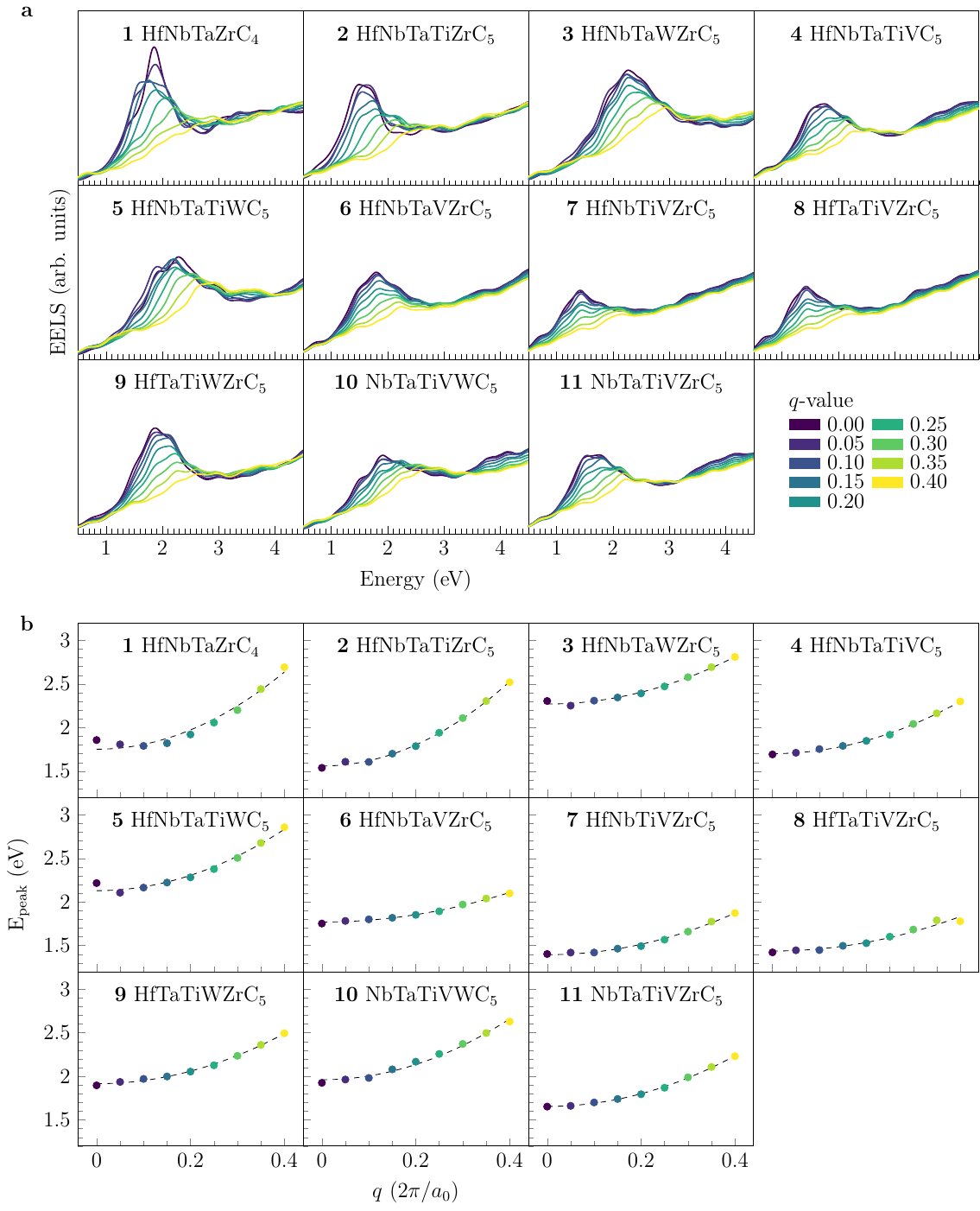}

 \caption{\small
 \textbf{\EELS\ spectra.}
 \textbf{a} Simulated \EELS\ for different values of $q$ (in units of $2\pi/\mathit{a}_0$ where $\mathit{a}_0$ is the lattice constant) for  eleven \HEC\ compositions.
 \textbf{b} Simulated dispersion $E_{\mathrm{peak}}(q)$ for  eleven \HEC\ compositions. The dashed lines are a guide for the eyes.}
 \label{fig:all_dispersion}
\end{figure*}

Quadratic dispersion relations ($E_0$ versus transferred momentum ${\bf
 q}$)
represent additional signatures of plasmonic excitations~\cite{Catellani_PRM_2020}.
Plasmon dispersions of our \PHECs\ were calculated with \DFT.
Given the computational cost of the entire \POCC\ ensemble, we limited it to the most probable \POCC\ tile (probability is the Boltzmann factor times the degeneracy; \ie\ the cardinality of the factor group.
The results for \RT\ are found to be quadratic and are shown in Figure~\ref{fig:all_dispersion}.
For our best plasmonic candidate \textbf{2}-\ch{HfNbTaTiZrC5}, we extend this analysis to the \HT\ regime ($T=$~\SI{1000}{\degreeCelsius}), also including the lattice expansion effect.
The results are shown in \textcolor{\Fcolor}{Figure \ref{fig:HfNbTaTiZrC5_Tdep}b}.
The corresponding $E({\bf q})$ curves still obey quadratic behavior, as in the \RT\ case.
These results confirm that the observed \EELS\ peaks in \HECs\ between \RT\ and \HT\ are a result of a plasmonic excitation.

\begin{figure}[ht!]
 \centerline{\includegraphics[width=0.5\textwidth]{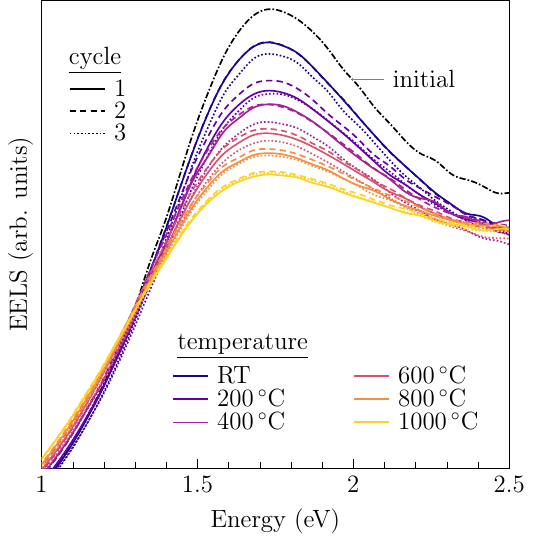}}
 \caption{\small \textbf{Temperature dependence of the plasmonic properties of \ch{HfNbTaTiZrC5}.}
 \EELS\ measured by ellipsometry over three heating-cooling cycles (\#1, \#2, and \#3).
Only spectra in the vicinity of the plasmonic resonance ($E_{\mathrm{peak}}$ $\approx$\ \SI{1.7}{eV}) are shown.
 }
 \label{fig:HfNbTaTiZrC5_heatcycles}
\end{figure}

\medskip
\noindent \textbf{Plasmonic resonance at variable temperatures} \\
 We tested the stability of the plasmonic resonance as a function of the temperature for system \textbf{2}:
three heating/cooling cycles (\RT$\rightleftarrows$\HT) are performed while optical properties are measured by ellipsometry.
The results are shown in \textcolor{\Fcolor}{Figure~\ref{fig:HfNbTaTiZrC5_heatcycles}} (a
magnification around the \EELS\ peak is shown in
\textcolor{\SIcolor}{\SFigure\ 8}).
The characteristics of the \EELS, as a function of the cycling, remain unchanged in the vicinity of the plasmon resonance, in sharp contrast to other plasmonic materials~\cite{Wells_OE_2018}.
In particular: {\bf i.} during every single heating/cooling process, the \EELS\ features change smoothly, without abrupt discontinuities; and {\bf ii.} the changes in plasmonic properties induced by heating return to the \RT\ state, indicating that the plasmonic properties of \PHECs\ are reversible with temperature over several cycles.
The reason lies in the phase stability of rock salt metal carbides. These compounds, and their binary precursors, are often sub-stoichiometric with {\it congruent melting} at few percent carbon-deficient from equicomposition. Their considerable range of solubility in the carbon-deficient region is bounded near equicomposition by a quite vertical rock salt $\leftrightarrow$ graphite {\it solvus}~\cite{Massalski}. Despite \HECs\ phase diagrams are not as well characterized as their binary and ternary carbide precursors, they are expected to follow the same high temperature trends~\cite{DEED}. This is quite advantageous: given a working temperature requirement and by playing with carbon composition, it is possible to synthesize \HECs\ immune from graphitic segregation during heat cycles (graphite is not plasmonically active in our range of interest~\cite{Djurisic_JAP_1999}), thus offering complete reversibility of the compound. Furthermore, low amount of oxygen impurities would not affect the plasmonic response as they would form transition metal oxycarbides with very low oxygen content at the invariant point~\cite{DEED}. In addition, carbon vacancies are not expected to disturb plasmonic resonance, which is dominated by the $d$-orbitals of the transition metals. The main effect of vacancies would be a change on the lifetime of the excitation given the interband transition of the carbon $p$-orbital electrons and \TM\ $d$-orbitals~\cite{Kumar_ACSPHOT_2016} (Figure 4a of Ref.~\onlinecite{curtarolo:art187} and Figure~\ref{fig:SI_DOS}).
While our testing temperatures are already much higher than the usual for optical applications, we expect plasmonic resonance to be preserved at even higher temperatures 1000-\SI{2000}{\degreeCelsius}, which are well below the melting temperature of \HECs~\cite{Liu_JECERS_2021}.

This implies that, at least for temperatures considered in this work, \PHEC\ samples remain almost structurally unchanged; \ie they do not undergo deformations that could affect bonding distribution/character and the overall free electronic charge density (responsible for the resonance).
The increased thermal vibrations cause the observed decrease and shift of plasmonic peak in the \EELS\ spectra through electron-phonon interactions.
Such thermal stability is a special feature of \PHECs\ over conventional plasmonic materials such as noble metals, which have lower melting points and would be irreversibly damaged at such temperatures.

\section*{\large Discussion}

\noindent In this article, we have experimentally verified that, depending on the composition, \HECs\ may have a plasmonic resonance both at \RT\ and \HT. We call this subset of carbides as \PHECs. We have also shown that such resonance is tunable with composition, and extremely stable under variable temperature conditions.
We found that excellent plasmonic stability and thermal cycling reversibility is achieved by designing the \PHECs\ with carbon composition inside the sub stoichiometric solubility range. Furthermore, the plasmonic resilience to oxygen impurities is ascribed to the existence of the oxicarbide invariant point at low oxygen concentration.

In summary, four remarkable facts arise from this work:
{\bf i.} the initial screening for good plasmonic \PHECs\ can be done
at \RT, avoiding \HT\ measurements that are cumbersome and expensive;
{\bf ii.} even though dissipative processes contribute to \EELS\ at \HT, they are not strong enough to dominate the plasmon resonance;
{\bf iii.} \PHECs\ that exhibit plasmonic response at \HT, maintain the same optical properties over the entire temperature range; and {\bf iv.} these \PHECs\ show considerable plasmonic thermal cycling stability.

Our findings reveal a new class of variable-temperature plasmonic high-entropy carbide materials. In particular, their optical properties, coupled with their well-known mechanical stability, make them potential promising prototypes for manufacturing thermal plasmonic technologies at variable temperatures.

\section*{\large Methods}

\noindent \textbf{Synthesis} \\
\noindent For the synthesis of \HEC\ pellets, individual carbide powder precursors were blended together in mixtures according to the targeted candidate compositions.
Starting powders consisted of
\ch{HfC} (99.0\%, H.C. Starck),
\ch{NbC} (99.5\%, Inframat Advanced Materials),
\ch{TaC} (99.5\%, Stanford Advanced Materials),
\ch{TiC} (99.7\%, Inframat Advanced Materials),
\ch{VC} (99.5\%, SkySpring Nanomaterials),
\ch{WC} (99.9\%, Inframat Advanced Materials),
\ch{ZrC} (99.9\% purity, U.S. Research Nanomaterials), and
graphite (99.0\%, Asbury Graphite Mills).
Batches were mixed by dry milling for 24 hours at \SI{200}{rpm} in a \SI{125}{mL} high-density polyethylene bottle; milling was performed in a 1:11 mass ratio of powder to \SI{10}{mm} {y}ttria-{s}tabilized {z}irconia (YSZ) media.
\FAST\ was used to densify the blended powders into \SI{20}{mm} bulk samples.
Sintering was carried out under $\sim$\SI{3}{mTorr} vacuum in a 25 Ton \FAST\ system (FCT Systeme GmbH),
with soak temperatures ranging between 2100 and \SI{2300}{\degreeCelsius}, soak times between 30 and \SI{60}{\minute},
and a ramped loading segment in which the applied pressure increased from 30 to \SI{50}{MPa} during the heating sequence (\SI{100}{\degreeCelsius/\minute}).
More details regarding this \FAST\ system setup for material
processing can be found in our previous reference~\cite{DEED}.

\medskip
\noindent \textbf{Ellipsometry} \\
\noindent To measure temperature-dependent spectroscopic ellipsometry, \HEC\ samples were polished to mirror finish and cleaned sequentially with acetone, ethanol, and deionized water.
Samples were then mounted individually on a sample holder using spot-welded tantalum strips.
After cleaning and mounting, samples were loaded into the load lock of a {m}olecular {b}eam {e}pitaxy (\MBE) system and, after reaching a base pressure lower than \SI{1e-6}{Torr}, outgassed at \SI{120}{\degreeCelsius} for \SI{60}{\minute}.
Upon finish outgassing and allowing the load lock to cool down overnight, samples were transferred into a model R450 \MBE\ reactor from DCA Instruments at a background pressure lower than \SI{5e-8}{Torr}.
In the \MBE\ reactor, \textit{in-situ} temperature-dependent ellipsometry measurements were performed using an M-2000 \textit{in-situ} ellipsometer model X-210 with near infrared upgrade from J.A. Woollam.
Single spectra were recorded in the range from 210 to \SI{1690}{nm} and acquired through the long spectrum acquisition mode in the J.A. Woollam CompleteEase software.
Sample height and alignment were calibrated to maximize ellipsometry signal intensity.
For each sample, an ellipsometry measurement was taken immediately after transfer into the \MBE\ chamber at room temperature.
All samples were then heated in sequence to \SI{1000}{\degreeCelsius} and cooled to 800, 600, 400, and \SI{200}{\degreeCelsius} using a ramp rate of \SI{30}{\degreeCelsius/\minute}.
After stabilizing at each of the target temperatures for \SI{5}{\minute}, ellipsometry spectra were recorded at background pressures below \SI{5e-8}{Torr}.
A final ellipsometry spectrum was taken in analogy to the initial measurement at room temperature when the sample temperature reached \SI{27}{\degreeCelsius} after the heating cycle.
Temperatures were measured by a thermocouple mounted inside the \MBE\ sample stage facing the backside of the sample.

\begin{figure}[ht!]
 \centerline{\includegraphics[width=0.5\textwidth] {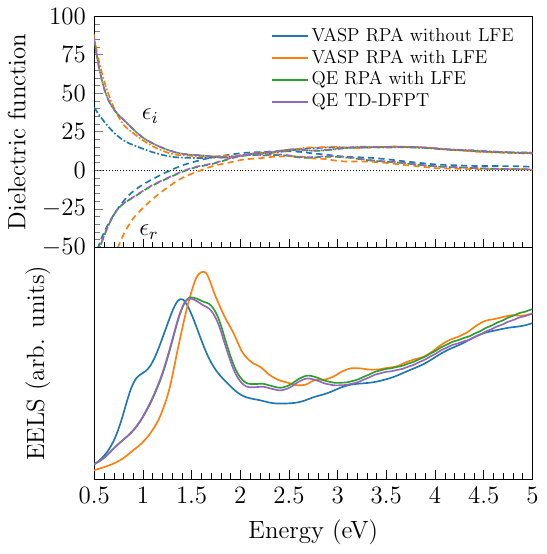}}
 \caption{\small \textbf{Comparison of ab-initio approximations.} Comparison of dielectric function and \EELS\ using \VASP\ with \RPA\ with and without \LFE, \QE\ with \RPA\ with \LFE, and \QE\ with time-dependent density functional perturbation theory (\TDDFPT) for a representative \POCC\ tile of \ch{HfNbTaTiZrC5}.}
 \label{fig:dft_method_comparison}
\end{figure}

\medskip
\noindent \textbf{Modeling the dielectric function} \\
\noindent Disordered systems are simulated by using the \POCC\ method, which considers the system as an ensemble average of ordered representative states --- tiles~\cite{aflowPOCC}.
For each tile $\mathit{j}$, \DFT\ is used to compute the factor-group degeneracy $g_j$, enthalpy $H_j$, and complex dielectric function $\hat{\epsilon}_j({\mathbf{q}},E)$.
Then, the complex dielectric function of the disordered system is given by:
\begin{equation}
 \label{eqn:df_avg}
 \langle \hat{\epsilon}_j({\mathbf{q}},E;T) \rangle = \sum_j P_j(T) \hat{\epsilon}_j({\mathbf{q}},E)
\end{equation}
where
$ P_j(T) = {g_j e^{-H_j/k_BT}}/{\sum_jg_j e^{-H_j/k_BT}}$
with $T$ and $k_B$ the experimental sintering temperature and the Boltzmann constant, respectively.
Unless otherwise stated, $\hat{\epsilon}_j$ is calculated in the optical limit $\mathbf{q=0}$,
using the Green-Kubo formula with the \VASP\ package, and {r}andom {p}hase {a}pproximation (\RPA) excluding {l}ocal-{f}ield {e}ffects (\LFE), \ie contributions due to the electron density inhomogeneity~\cite{Gajdos2006}.
To handle the case when $\mathbf{q}\neq\mathbf{0}$, we use Quantum Espresso (\QE) in the \RPA\ including \LFE~\cite{Giannozzi:2017io}.
All calculations are performed using the {P}erdew-{B}urke-{E}rnzerhof
functional~\cite{PBE} with the convergence parameters set
(energy cutoff, k-point sampling, etc.) by the \AFLOW\ standard~\cite{curtarolo:art191}.
In Figure~\ref{fig:dft_method_comparison},
we show the complex dielectric function and \EELS\ spectra for a representative \POCC\ tile of \ch{HfNbTaTiZrC5}, calculated using the Drude-Lorentz model
within the random phase approximation (\RPA), as implemented in \VASP\ and \QE\ codes, with and without the local field effects (\LFE).
The results are also compared with spectra simulated by using the time-dependent density functional perturbation theory (\TDDFPT)
approach implemented in the {\em turbo\_eels} code, also included in the \QE\ suite.
In the latter case, spectra are evaluated in the limit of vanishing transferred momentum (\textbf{q$\rightarrow$0}).
The four set of spectra are in excellent agreement, within the numerical errors.
The small deviation between the \VASP\ and \QE\ \RPA\ results is attributed to local field effects (\LFE), which can slightly shift the peak positions, especially at low energies~\cite{Hanke_PRB_1975}.

\medskip
\noindent \textbf{Data availability} \\
\noindent The code used to generate and process dielectric simulation data will be publicly available upon the release of the next version of \AFLOW.
Processed data used in this work are also available from the corresponding author upon reasonable request.

\medskip
\noindent \textbf{Acknowledgements} \\
\noindent The authors thank
Cormac Toher,
Scott Thiel,
Michael Mehl, and
Lucas Wilson
for fruitful discussions.
The authors also thank Stephanie Law and Maria Hilse for providing assistance with the ellipsometry experiments.
This research was supported by the Office of Naval Research under grants N00014-23-1-2615 and N00014-24-1-2768.
This work was supported by high-performance computer time and resources from the DoD High Performance Computing Modernization Program (Frontier).
We acknowledge Auro Scientific, LLC for computational support.

\medskip
\noindent \textbf{Contributions} \\
\noindent
Authors Simon Divilov, Sean Griesemer, and Robert Koennecker contributed equally to the project.
S.~C. and A.~C. conceptualized the disorder expansion for dielectric properties.
S.~D. and S.~D.~G. performed the simulations and wrote the code to process simulation results.
R.~C.~K. and M.~J.~A. synthesized the samples, performed experiments and contributed to writing the paper.
A.~C.~Z. co-wrote the code to process simulation results.
H.~E., S.~D. and S.~C. wrote appropriate parts of the \AFLOW\ code and advised on the simulations.
J.~R.~S. advised on the \REELS\ technique.
X.~C. contributed to paper writing.
A.~C., D.~E.~W., and S.~C. established the research direction and supervised the project.
All the authors contributed to the writing of the manuscript.

\medskip
\noindent \textbf{Conflict of interest} \\
\noindent The authors declare no competing interests.

\medskip
\noindent {\textbf{Supplementary information} \\
The online version contains supplementary material available at...

\medskip

\newcommand{\Ozolins}{Ozoli{\c{n}}{\v{s}}}

\end{document}